\documentclass[12pt]{spieman}  % 12pt font required by SPIE;
\usepackage{amsmath,amsfonts,amssymb}
\usepackage{graphicx}
\usepackage{setspace}
\usepackage{tocloft}
\usepackage{lineno}
\usepackage{multirow}
\usepackage{multicol}
\usepackage{wrapfig}

\title{Ultraviolet observations of atmospheric escape in exoplanets with the \emph{Habitable Worlds Observatory}}

\author[a, b, *]{Leonardo A. Dos Santos}
\author[c, d]{Eric Lopez}
\affil[a]{Space Telescope Science Institute, 3700 San Martin Drive, Baltimore, MD 21218, USA}
\affil[b]{Department of Physics and Astronomy, Johns Hopkins University, 3400 N. Charles Street, Baltimore, MD 21218, USA}
\affil[c]{NASA Goddard Space Flight Center, 8800 Greenbelt Rd, Greenbelt, MD 20771, USA}
\affil[d]{GSFC Sellers Exoplanet Environments Collaboration, NASA GSFC, Greenbelt, MD 20771, USA}

\cftpagenumbersoff{figure}
\cftpagenumbersoff{table} 
\begin{document} 
\maketitle

\begin{abstract}
Among the many recommendations of the Decadal Survey on Astronomy and Astrophysics 2020, we found that a priority area of research is to pave the pathways towards finding and characterizing habitable worlds. In this context, we aim to understand how planetary systems evolve through atmospheric escape, and develop techniques to identify potentially Earth-like worlds. Using the ultraviolet (UV) capabilities of the \emph{Habitable Worlds Observatory}, we can use transit spectroscopy observations to determine what processes drive the evolution of exoplanets, how well can small exoplanets retain atmospheres, and search for Earth-like atmospheres. We advocate the development of a UV spectrograph that is capable of moderate- to high-resolution spectroscopy of point sources, access to key spectral features between 1000 and 3000~\AA, and UV detectors that are resilient to high count rates.
\end{abstract}

% Include a list of up to six keywords after the abstract
\keywords{ultraviolet, spectroscopy, exoplanets, habitability, atmospheres}

% Include email contact information for corresponding author
{\noindent \footnotesize\textbf{*}Leonardo A. Dos Santos, \linkable{ldsantos@stsci.edu}}

\begin{spacing}{2}   % use double spacing for rest of manuscript

\section{Introduction}
\label{sect:intro}  % \label{} allows reference to this section

The discovery of 51~Peg~b, a short-period, hot gas giant planet outside of the Solar System\cite{Mayor1995} ushered the development of a whole new sub-field of astronomy now informally known as exoplanet science. This discovery was in strong tension with our knowledge of how planets form and evolve: how can a hot Jupiter end up orbiting so close to a Sun-like star without completely losing its H+He envelope due to evaporation? The first attempts to explain this conundrum concluded that not only can hot Jupiters survive evaporation\cite{Guillot1996}, but signals of their evolution would be imprinted in their spectra\cite{Seager1998}. Hot Jupiters like 51~Peg~b are cool enough that chemistry plays an important role in determining the composition and structure of their photosphere. On the other hand, their upper atmospheres see a sharp increase in temperature due to the absorption of high-energy photons from the host star. This increase in temperature causes the onset of an expanding outward wind that drives atmospheric escape rates high enough to affect how small planets evolve, a process known as hydrodynamic escape\cite{Watson1981}.

Atmospheric escape has been studied in the Solar System planets since the early 20th Century\cite{Jeans1925, Spitzer1949, Opik1963, Chamberlain1963}. However, the modern planets do not offer us a clear window in to hydrodynamic escape, as most of them remain in the Jeans regime\cite{Jeans1925} of evaporation -- except for Pluto\cite{Tian2005}. Hydrodynamic escape was originally proposed to explain the loss of H/He envelopes in the early Earth and Venus\cite{Watson1981}, but the evaporation of H was only observed for the first time in the exoplanet HD~209458b\cite{Vidal2003}. The authors led an observing program with the \emph{Hubble Space Telescope} (\emph{HST}) that leveraged the transmission spectroscopy technique to detect atmospheres of transiting exoplanets\cite{Seager2000, Charbonneau2002}. See Section \ref{sec:obs} for more details on the technique and previous observations. See Gronoff et al. \cite{Gronoff2020} and Hazra et al. \cite{Hazra2025} for detailed reviews on the different physical mechanisms of atmospheric escape and its impact in planetary evolution. 

With the launch of the Kepler satellite and the resulting statistical sample of transiting exoplanets with measured radii\cite{Batalha2013}, we had the first evidence for the impact of hydrodynamic escape in the evolution of sub-Jovian exoplanets. There are two demographic features in particular that are deemed to provide such evidence: the hot-Neptune desert\cite{Szabo2011, Mazeh2016, Owen2018} and the super-Earth radius valley\cite{Owen2017, Fulton2017}. Both these features exhibit a dearth of planets within a range of radii that we hypothesize to be carved by intense and rapid atmospheric escape due to high-energy irradiation, also known as photoevaporation \cite{Owen2013}. Both the direct observation of hydrodynamic escape in transiting exoplanets and the indirect detection of the effects of photoevaporation in exoplanet demographics provide a first look into the evolution of exoplanets. However, many questions still remain open.

\begin{itemize}
    \item What is the efficiency of converting high-energy stellar photons into an outflow?
    \item What is the chemical composition of planetary outflows and how does it affect mass-loss rates?
    \item What is the role of magnetic fields in enhancing or suppressing atmospheric escape?
    \item What are the timescales of mass loss in small, potentially Earth-like exoplanets?
    \item Do planets outside of the Solar System have exospheres similar to the Earth's?
\end{itemize}

In order to answer these questions, we need an ultraviolet (UV) instrument capable of performing precise, high spectral resolution, uninterrupted time series observations of exoplanet transits. Although \emph{HST} and ground-based spectrographs have been the workhorses of atmospheric escape observations thus far, a large swath of the parameter space remains unexplored. In this context, the Habitable Worlds Observatory (HWO) not only is posed to kickstart the search for life outside the Solar System, but it will be revolutionary for studies of exoplanet astrophysics. 

\section{State-of-the-art observations}\label{sec:obs}

The main technique used to observe atmospheric escape in exoplanets is called transmission spectroscopy \cite{Seager2000, Brown2001, Hubbard2001}. Qualitatively, it consists on observing the spectrum of a star that is being transited by an exoplanet and measuring changes in the spectra that indicate the presence of an extra source of opacity, presumably the planet's atmosphere. 

\subsection{A primer on transmission spectroscopy}\label{sec:tech}

In its simplest form, the transmission spectrum $\Phi$ is given by:

\begin{equation}
    \Phi(\lambda) = 1 - \frac{F_\mathrm{in}(\lambda)}{F_\mathrm{out}(\lambda)} \mathrm{,}
\end{equation}where $F_\mathrm{out}$ is the out-of-transit spectrum of the host star and $F_\mathrm{in}$ is the observed in-transit spectrum. At high enough spectral resolution, the planetary signal can be resolved, and the wavelength of the excess absorption will have a dependence on transit phase. In this case, the transmission spectrum $\phi$\footnote{\footnotesize{The lower case denotes the dependence of the in-transit spectra to the orbital phase $\theta$.}} as a function of the planetary phase $\theta$ and wavelength $\lambda$ is the following:

\begin{equation}
    \phi(\theta, \lambda) = 1 - \frac{f_\mathrm{in}(\theta,\lambda)}{F_\mathrm{out}(\lambda)} \mathrm{.}
\end{equation}We can also define the transmission spectrum in the rest frame of the planet by Doppler shifting the spectrum according to:

\begin{equation}
    \lambda_\mathrm{p}(\theta) = \lambda \left(\frac{\Delta v(\theta)}{c} + 1 \right) \mathrm{,}
\end{equation}where $\Delta v$ is the difference between radial velocity of the planet at a particular phase $\theta$ and the reference velocity of the observer, and $\lambda_\mathrm{p}$ will be the resulting wavelength in the rest frame of the planet. Finally, we can define the transit-integrated transmission spectrum $\Phi$ by taking the mean of $\phi(\theta, \lambda_\mathrm{p})$ over the range of $\theta$ observed in transit:

\begin{equation}
    \Phi (\lambda_\mathrm{p}) = \frac{1}{\Delta \theta}\int \phi(\theta, \lambda_\mathrm{p}) d\theta \mathrm{.}
\end{equation}

Upper atmospheres are populated mostly by atomic and ion species, which in turn tend to absorb light more efficiently in individual resonance lines in the optical and ultraviolet (UV) wavelengths. This is in contrast with the lower atmosphere, which is dominated by molecular species and the bulk of the absorption of light in transmission takes place in the continuum and in the forest of molecular features in near-infrared (NIR) and longer wavelengths. There are notable exceptions to this general rule, such as the metastable He line at 1.083~$\mu$m, which traces atmospheric escape \cite{Oklopcic2018}, and the forest of lines of some molecular species in the near-UV \cite{Lothringer2020}.

Since the signals of atmospheric escape usually take place in individual spectral features, a significant fraction of information is contained in the shape of the absorption that planets imprint in the spectra of their stars as they transit. Furthermore, the Doppler velocity information contained in time-series of transmission spectra are routinely used to study atmospheric dynamics \cite{Bourrier2016, Ehrenreich2020} and disentangle stellar contamination from planetary signals \cite{Casasayas2021}. In fact, some of these signals can be completely undetectable at low spectral resolution \cite{DSantos2023b}.

\subsection{Current capabilities with \emph{HST}}

\begin{figure}
\begin{center}
\begin{tabular}{c}
\includegraphics[height=9.5cm]{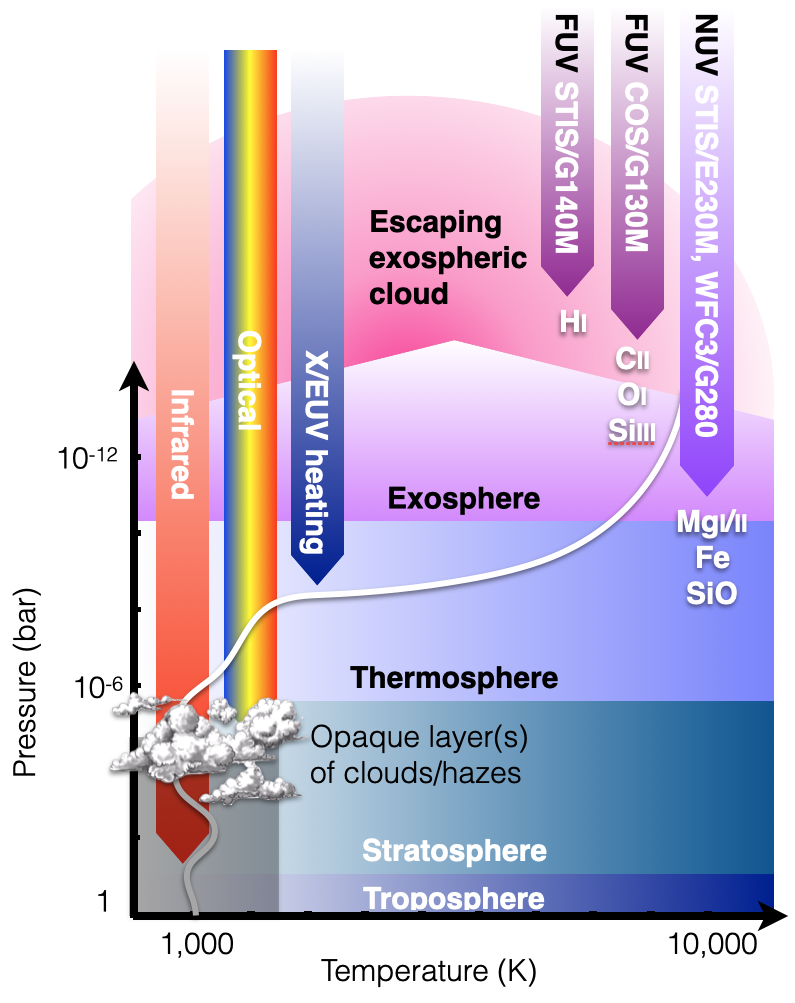}
\end{tabular}
\end{center}
\caption 
[Schematic representation of an exoplanet atmosphere.]{ \label{fig:structure}
Schematic representation of an exoplanet atmosphere and pressure-temperature profile. Different wavelengths are absorbed at different pressure levels; hence they can be combined to probe the whole atmospheric structure \cite{VMadjar2011}. Here, we focus on the upper atmosphere: the thermosphere and the exosphere. Credit: D. Ehrenreich.} 
\end{figure}

What makes \emph{HST} unique for exoplanet observations is its access to the ultraviolet (UV; see Table \ref{tab:hst} for a summary of the relevant observing modes and their characteristics\footnote{\footnotesize{More detailed information can be found in the corresponding Instrument Handbooks available at \url{https://hst-docs.stsci.edu/}.}}). At these wavelengths, we are sensitive to the upper layers of planetary atmospheres, between the thermosphere and the exopshere (see Fig. \ref{fig:structure}), where hydrostatic equilibrium does not apply anymore. Due to the low pressures and densities, material at these layers is strongly affected by several thermal and non-thermal mechanisms, such as atmospheric escape \cite{Tian2013}, electric and magnetic interactions \cite{Shizgal1996, Khodachenko2021}, high-altitude winds \cite{Seidel2019}, tidal effects \cite{Lecavelier2004}, photochemistry \cite{Zhao2015} and impacts \cite{Haff1981, Schlichting2015}. Thus, UV presents a rich window into the many different physical processes that dictate the evolution of exoplanet atmospheres.

\begin{table}[ht]
    \centering
    \caption{UV observing modes available with \emph{HST} relevant for exoplanet observations.}
    \begin{tabular}{clccc}
    \hline
        Instrument & Optical element & \multicolumn{2}{c}{Wavelength range [\AA]} & Resolving power \\
         &  & (complete) & ($\Delta$ per tilt) & ($R$, approximate) \\
        \hline
       STIS  & G140L & 1150–1730 & - & 1,000 \\
         & G230L & 1570–3180 & - & 500 \\
         & G140M & 1140–1740 & 55 & 10,000 \\
         & G230M & 1640–3100 & 90 & 10,000 \\
         & E140M & 1144–1729 & 567 & 46,000 \\
         & E230M & 1607–3129 & 800 & 30,000 \\
         & E140H & 1140–1700 & 210 & 114,000 \\
         & E230H & 1626–3159 & 267 & 114,000 \\
         \hline
      COS & G140L & 900-2150 & 1133 & 2,000 \\
      & G230L & 1650-3200 & 3$\times$400 & 3,000 \\
         & G130M & 900-1450 & 2$\times$140 & 14,000 \\
         & G160M & 1360-1775 & 2$\times$170 & 16,000 \\
         & G185M & 1700-2100 & 3$\times$35 & 18,000 \\
         & G225M & 2100-2500 & 3$\times$33 & 22,000 \\
         \hline
      WFC3 & G280 & 2000-8000 & - & 350 \\
      \hline
    \end{tabular}
    \label{tab:hst}
\end{table}

% \end{figure}

The first detection of atmospheric escape in an exoplanet was performed with \emph{HST} and the Space Telescope Imaging Spectrograph (STIS) \cite{Vidal2003}. It consisted in a transit spectra time-series of the Lyman-$\alpha$ line (1215.67~\AA), which exhibited an absorption of $15\%$ at high velocities, blueshifted beyond 100~km\,s$^{-1}$. This striking signal cannot be explained by a simplified one-dimensional hydrodynamical flow from the planet's upper atmosphere \cite{MClay2009}; instead, it requires more sophisticated models that include acceleration by radiation pressure and interactions with the stellar wind \cite{Bourrier2013} to more accurately fit the observation. If the spectral resolution of STIS was not high enough to resolve the Lyman-$\alpha$ absorption in velocity space, this valuable information would have been lost, misinterpreted or even undetectable. The discovery discussed above opened the floodgates for a wave of theoretical efforts to study exospheric dynamics in transiting exoplanets \cite{Holmstrom2008, Lecavelier2008, Vidotto2017, Wang2018, Debrecht2020}. Even non-detections in Lyman-$\alpha$ observations with \emph{HST} have provided us with constraints on the interactions between stellar and planetary winds, as demonstrated in Vidotto \& Cleary \cite{Vidotto2020}.

% Metals
Neutral H is not the only tracer of atmospheric escape in the UV and, in fact, these wavelengths are the richest for observing escaping metallic species, such as C, N, O, Si in the far-ultraviolet (FUV) \cite{Vidal2004, DSantos2019b} and Mg, Fe, and SiO in the near-ultraviolet (NUV) \cite{Sing2019, Lothringer2022}. Observing these species is crucial because they are a direct tracer of hydrodynamic escape, where the mass-loss rates are so intense that the outflowing H drags up heavier atoms towards the upper atmosphere -- this process is the most important ingredient in evolution of planetary atmospheres, even those that are Earth-like \cite{Watson1981, Zahnle2017}.

% Small exoplanets in the UV
UV observations with \emph{HST} have enabled us to detect atmospheres in exoplanets that even \emph{JWST} does not have access to. For example, transiting exoplanets with high metallicity or cloudy atmospheres are challenging to characterize in the optical or NIR \cite{Kreidberg2014, Alam2025, Mukherjee2025}. Since the UV traces upper layers of their atmospheres, clouds and metallicity are not an issue and may in fact be favorable for the detection of escaping metals. The first detection of an atmosphere in a super-Earth, $\pi$~Men~c, was reported in Garc\'ia Mu\~noz et al. \cite{GMunoz2021}, where the investigators observed escaping C in the planet's \emph{HST}/STIS transmission spectrum. The cloudy warm Neptune GJ~436b, whose atmosphere is seen as a flat spectrum in the optical and NIR \cite{Pont2009, Lothringer2018}, exhibits the strongest atmospheric signal ever detected for an exoplanet in Lyman-$\alpha$ \cite{Ehrenreich2015, Lavie2017, DSantos2019b}. Recent \emph{HST} UV observations have also unlocked the atmospheric properties of the sub-Neptune system in TOI-776 \cite{Loyd2025}, whose \emph{JWST} observations had previously rendered only a flat transmission spectrum \cite{Alderson2025}.

Another recent development in the field is the usage of \emph{HST}'s Wide-Field Camera 3 (WFC3) in the near-UV, through the G280 grism, to observe exoplanet atmospheres. Using this mode, Wakeford et al. \cite{Wakeford2020} measured the first transmission spectrum with WFC3/G280 for the hot Jupiter HAT-P-41b, which revealed a high-metallicity, cloudy atmosphere. Similarly, Lothringer et al. \cite{Lothringer2022} measured an increase in the opacity at NUV that is best explained by the presence of metals and SiO in the atmosphere of the ultra-hot Jupiter WASP-178b. Unfortunately, the authors were unable to identify exactly what these metal species are or whether they are escaping due to the low spectral resolution of WFC3; albeit they attempted to identify the metals using STIS high-resolution spectroscopy in the NUV, the lower sensitivity of the {\it \'echelle} mode precluded a firm detection.

We refer the reader to sections 3 and 4 of the review in Dos Santos \cite{DSantos2023a} for a summary of all the UV atmospheric escape observations in exoplanets up to the year 2022. More recent UV observations with \emph{HST} and the Colorado Ultraviolet Transiting Experiment (CUTE) have been reported in the literature as well \cite{DSantos2023c, Rockcliffe2023, Sreejith2023, Egan2024, Morrissey2024, Loyd2025}.

\subsection{Current capabilities with other facilities}

Although the first atmospheric escape observations were conducted from space with \emph{HST}, ground-based observatories have also been productive in observing this physical process in transiting exoplanets. The first efforts were focused on detecting upper-atmospheric H using H$\alpha$ transmission spectroscopy \cite{Jensen2012, Cauley2017}. More recently, the discovery of the metastable He line at 1.083~$\mu$m as a tracer of escape \cite{Seager2000, Oklopcic2018} has opened the doors for a wave of ground-based surveys in gas giants and sub-Jovian worlds \cite{DSantos2023a, Orell2024, McCreery2025}.

Crucial to these observations is the high spectral resolution allowed by optical and near-IR, ground-based spectrographs (see more details in Section \ref{sec:resolution}). But, in some cases, narrow-band photometry has been used to detect escaping He \cite{Vissapragada2020, Levine2024}, and low-resolution spectroscopy with the \emph{James Webb Space Telescope} (\emph{JWST}) \cite{Fu2022, DSantos2023b}.

The main challenge of these tracers is that they are not resonant lines, but rather lines that depend on a fine-tuned cascade of ionization, recombination, (de-)excitation and collisional processes in order to produce a detectable signal \cite{Shaikhislamov2024, Allan2024}. This limits the types of planets and host stars that we can probe to specific cases, such as ultra-hot Jupiters \cite{Wyttenbach2020, Seidel2025} and stars with relatively hard spectra \cite{Oklopcic2019}.

\section{Capabilities of UV transit spectroscopy with \emph{HWO}}

Previous concept studies of the Large Ultraviolet-Optical-Infrared Surveyor (LUVOIR) \cite{Bolcar2017a, Bolcar2017b, France2017} and the Habitable Exoplanet Observatory (HabEx) \cite{Gaudi2018} were fundamental to identify the necessary capabilities of the next NASA flagship mission to revolutionize astrophysics and exoplanet science. Most recently, the Astro2020 Decadal Report \cite{NAP26141} promoted the development of \emph{HWO}, which is posed to be a more powerful successor to \emph{HST}.

\subsection{Desirable technical aspects}

The exact design and capabilities of \emph{HWO} have not been decided yet, but we shall discuss the required and desired technical aspects of the telescope and its UV spectrograph in order to explore the next frontier of research in atmospheric escape and planetary evolution.

\subsubsection{Spectral resolving power}\label{sec:resolution}

One of the most important lessons learned in previous observations of atmospheric escape, both from space and from the ground, is that moderate to high spectral resolution is an absolute must. It is needed not only to disentangle the planetary signal from stellar contamination \cite{Casasayas2021}, but also for its theoretical interpretation using atmospheric escape models \cite{Lampon2021, Vissapragada2022, DSantos2023a}.

Simulations of transmission spectra for atmospheric escape observations have revealed that planetary upper atmospheres imprint absorption features with a width in the order of tens of km\,s$^{-1}$ for spherically symmetric outflows \cite{DSantos2022}, which is broadened by a combination of temperature, vertical winds and rotation. For Earth-like exoplanets, since the escape is in Jeans regime, we expect atomic species to be near escape velocity ($\sim$11.2~km\,s$^{-1}$ for Earth's gravity), but they can be accelerated to higher velocities due to radiation pressure and interactions with the stellar wind. In order to resolve this absorption, a bare minimum spectral resolving power of $R \sim 10\,000$ is required. This is similar to the capabilities currently available with \emph{HST}'s COS and STIS spectrographs in their first-order M modes (see Table \ref{tab:hst}). As seen in previous observations, some escape signals can depart from spherical symmetry and absorb stellar light at hundreds of km\,s$^{-1}$ in the stellar rest frame due to non-thermal effects \cite{Vidal2003,Ehrenreich2015, Loyd2025}.

Further transformational research will, however, be enabled if we have access to high spectral resolving power similar to or better than the E gratings of STIS ($R > 40\,000$), but without the sensitivity losses of an {\it \'echelle} cross-disperser (see Section \ref{sec:sens}). At these levels of resolving power, a given transiting planet imprints a Doppler movement over the stellar atmosphere in the order of a few km\,s$^{-1}$ \cite{Cegla2023}, which allows the observer to confirm the planetary nature of the signal \cite{Allart2019} and disentangle it from the Rossiter-McLaughlin effect \cite{Cegla2016}.

\subsubsection{Wavelength range}

\begin{figure}[t]
    \begin{center}
    \begin{tabular}{c}
    \includegraphics[width=0.97\textwidth]{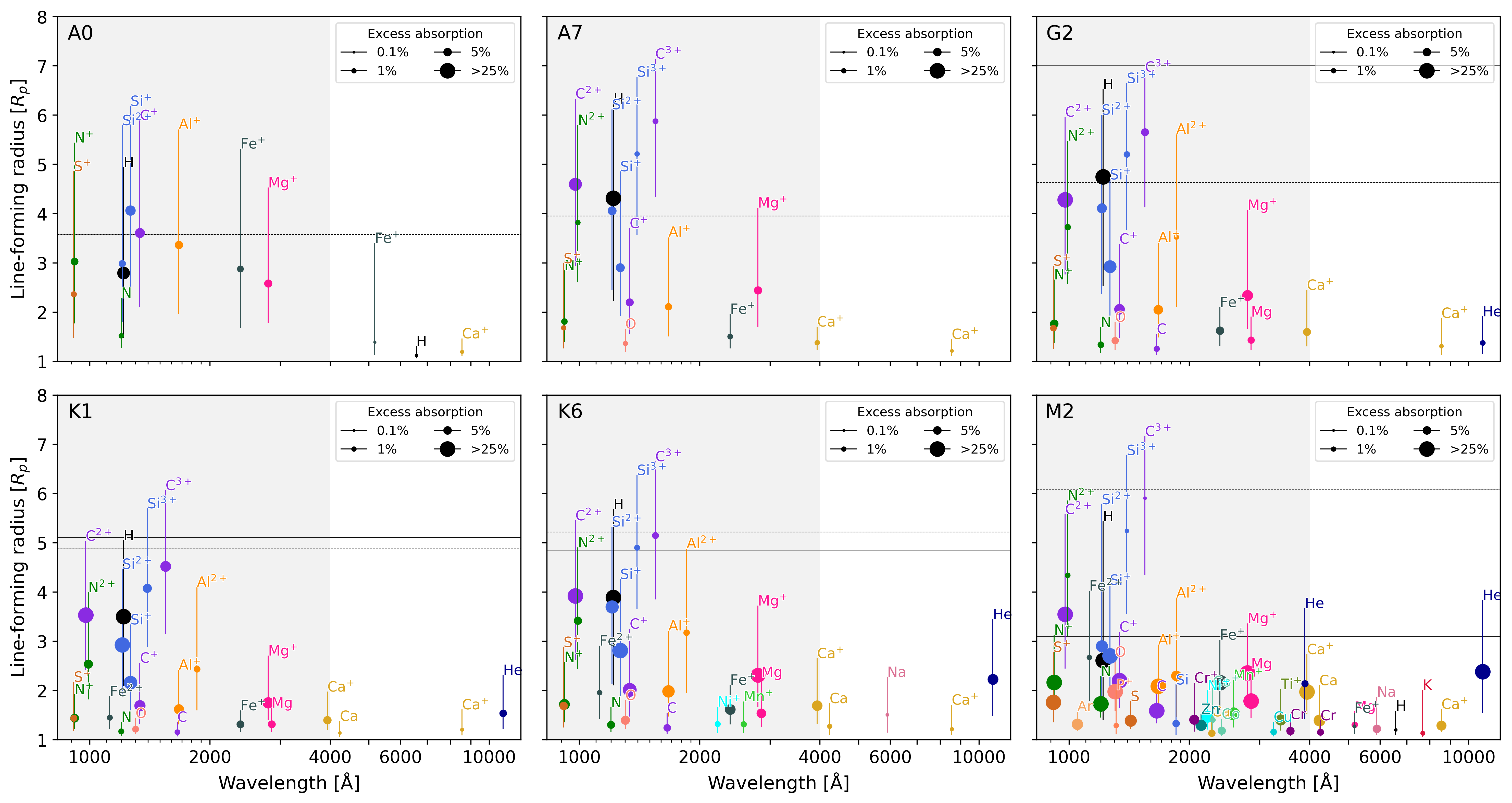}
    \end{tabular}
    \end{center}
    \caption 
    [Spectral features that trace atmospheric escape.]{\label{fig:lines}
    Spectral features detectable in transmission spectra of hot Jupiters as a function of wavelength and stellar host type \cite{Linssen2023}. The wavelengths relevant for UV spectroscopy are those below 3000~\AA. The line-forming radius (y-axis) is a proxy for the altitude where the feature is located in relation to the planetary surface and the symbol sizes represent the depth of the excess absorption in transmission. \copyright D.~C.~Linssen \& A. Oklop{\v{c}}i{\'c}, used with permission.} 
    \end{figure}

The UV spectrum is rich in individual lines that trace atmospheric escape in transiting exoplanets. Most detections to date have been obtained in the Lyman-$\alpha$ line at 1215.67~\AA\ and the Mg features at 2796.35, 2803.53 and 2852.965~\AA. A recent investigation by Linssen et al. \cite{Linssen2023} provided a thorough, model-based compilation of lines that trace escape. In the UV, these lines are contained within the range 970 to 3000~\AA, where the densest cluster of features lies within 1000 to 2000~\AA\ (see Fig. \ref{fig:lines}). Linssen et al. \cite{Linssen2023} further predicted that, after Lyman-$\alpha$, the strongest absorber lies in the C\,{\sc iii} at 977~\AA, which is currently accessible to \emph{HST}/COS, albeit at an extremely low sensitivity. According to them, combining this C\,{\sc iii} with other C lines at varying levels of ionization within 1300 to 1700~\AA\ would help better constrain the temperature and velocity profiles of exoplanet outflows.

\subsubsection{Detector sensitivity and dynamic range}\label{sec:sens}

\begin{figure}[t]
    \begin{center}
    \begin{tabular}{c}
    \includegraphics[width=0.75\textwidth]{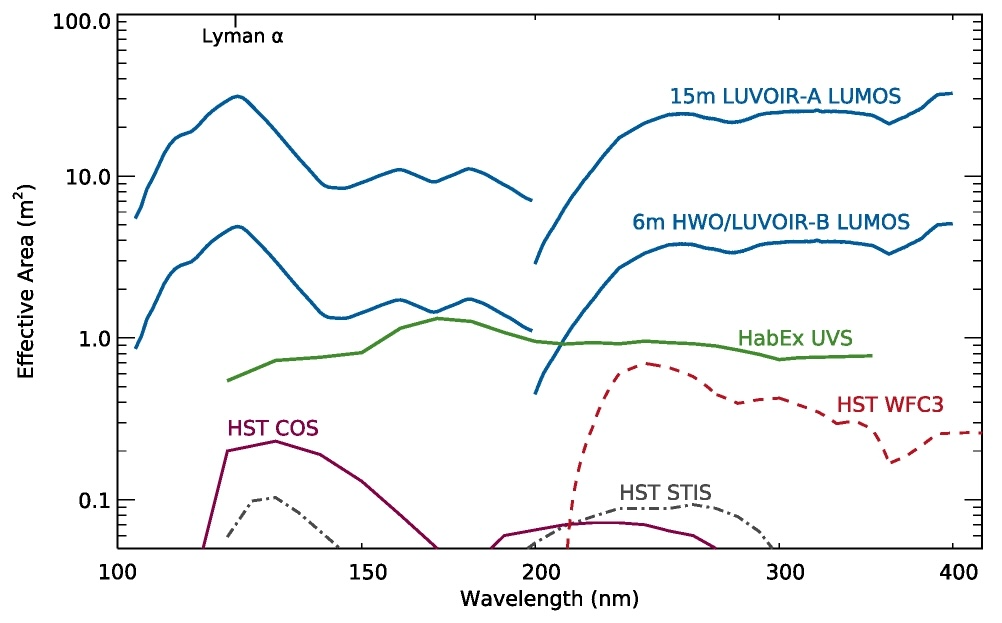}
    \end{tabular}
    \end{center}
    \caption 
    [Effective area of \emph{HST} and \emph{HWO} instruments.]{ \label{fig:effective_area}
    Effective area of different instruments capable of UV spectroscopy. According to its concept studies \cite{Bolcar2017b, France2017}, the \emph{LUVOIR}/LUMOS instrument would have an effective area that is more than ten times better than the COS and STIS instruments on \emph{HST}, assuming the more conservative telescope aperture size (LUVOIR-B).} 
    \end{figure}

For exoplanet UV observations, the truly transformational advance of \emph{HWO} lies within its significant improvement in sensitivity compared to \emph{HST}. The concept study for the LUVOIR Ultraviolet Multi-Object Spectrograph (LUMOS) provided a first look into this improvement \cite{France2017}, which also offer point-source capabilities. A combination of the LUMOS instrument with the LUVOIR-B design is predicted to offer $\sim 100 \times$ increased effective area compared to \emph{HST}/COS in the FUV and $\sim 10 \times$ improvement compared to \emph{HST}/WFC3 in the NUV (see Fig. \ref{fig:effective_area}).

As demonstrated in Dos Santos et al. \cite{DSantos2019a}, such an improved effective area will enable the search of Earth-like exospheres in exoplanets transiting M dwarfs (see also G\'omez de Castro et al. \cite{Castro2018}). The Earth's exosphere extends to more than 38~R$_\oplus$ and is composed mainly of neutral H atoms \cite{Kameda2017}. These atoms are the product of dissociation of water molecules in the lower atmosphere followed by Jeans escape, so finding Earth-like exospheres around extrasolar planets could be used as an indication of evaporated oceans or habitable conditions in the lower atmosphere \cite{Jura2004}. Currently, the capabilities of \emph{HST} are far below the requirements necessary to perform such a detection, so \emph{HWO} will be strictly required to observe Earth-like exospheres outside of the Solar System.

In order to achieve the high signal-to-noise ratios required to observe exospheres around rocky exoplanets, we will need to observe bright stars. One of the most important limitations of \emph{HST} in the UV is the low dynamic range of its UV detectors, the Multi-Anode Microchannel Arrays (MAMA) and the Cross-Delay Line (XDL) on COS and STIS. Due to their technical characteristics and design, these detectors cannot operate on count rates above $\sim 100$~cts\,s$^{-1}$\,px$^{-1}$ at the local level or $770\,000$~cts\,s$^{-1}$ at the global level in the NUV (in the FUV, the global limits are $60\,000$~cts\,s$^{-1}$ for COS and $120\,000$~cts\,s$^{-1}$ for STIS). Ideally, the UV detectors on \emph{HWO} should be designed to have a higher dynamic range, which would not only allow us to perform the observations described above, but also study stellar flares without the risk of damaging the detectors \cite{Loyd2018}.

\subsubsection{Telescope orbit}

The \emph{Hubble Space Telescope} has two important limitations related to its orbital motion that affect the quality of exoplanet observations in the UV: geocoronal contamination and Earth occultations. The low-Earth of \emph{HST} orbit sits within our planet's exosphere, which is rich in H, He and O atoms. These atoms scatter solar photons \cite{Kameda2017}, which hit the detectors in \emph{HST} and produce geocoronal contamination. For observations of planet-hosting stars, this contamination has similar or stronger amplitude than stellar emission and imposes significant, additional noise in the core of geocoronal lines because of the photon-counting nature of UV detectors \cite{Aguirre2023}.

The orbital motion of \emph{HST} also causes targets to be occulted by the Earth unless they are in its continuous vieweing zone near the orbital poles. In practice, this has two important implications: i) time-series observations have interruptions with a period of $\sim$95~minutes and durations that can last approximately half of the period; ii) the temperature variation as the telescope moves between day and night in its orbit causes physical changes that affect optical properties and affects the accuracy of flux calibration -- this is also known as ``breathing'' \cite{Bely1993}. Both these effects impose limits in the quality of time-series observations: for example, it is common to lose information about limb darkening if the observations happen to miss the transit ingress or egress, which translates into larger uncertainties or error in transit depth when fitting transit light curves. Furthermore, telescope breathing also requires modeling of this effect, which incurs in higher uncertainties and potential measurement errors \cite{Sing2019}.

Ideally, short-cadence time-series observations with \emph{HWO}, like exoplanet transits, would benefit from uninterrupted exposures, similar to what is achieved by \emph{JWST}. Furthermore, moving above the low-Earth orbit would minimize geocoronal contamination of H, O and N lines that are present in the UV. Thus, an orbit located at the L2 equilibrium point would achieve both objectives. 

\subsection{Future observing campaigns}

To take advantage of the new opportunities that \emph{HWO} will open for exoplanet science in the UV, we advocate the implementation of two observing programs that follow distinct approaches to answer the set of questions exposed in Section \ref{sect:intro}.

\subsubsection{A ``deep-field'' search for an Earth-like exosphere}

\begin{figure}[t]
    \begin{center}
    \begin{tabular}{c}
    \includegraphics[width=0.75\textwidth]{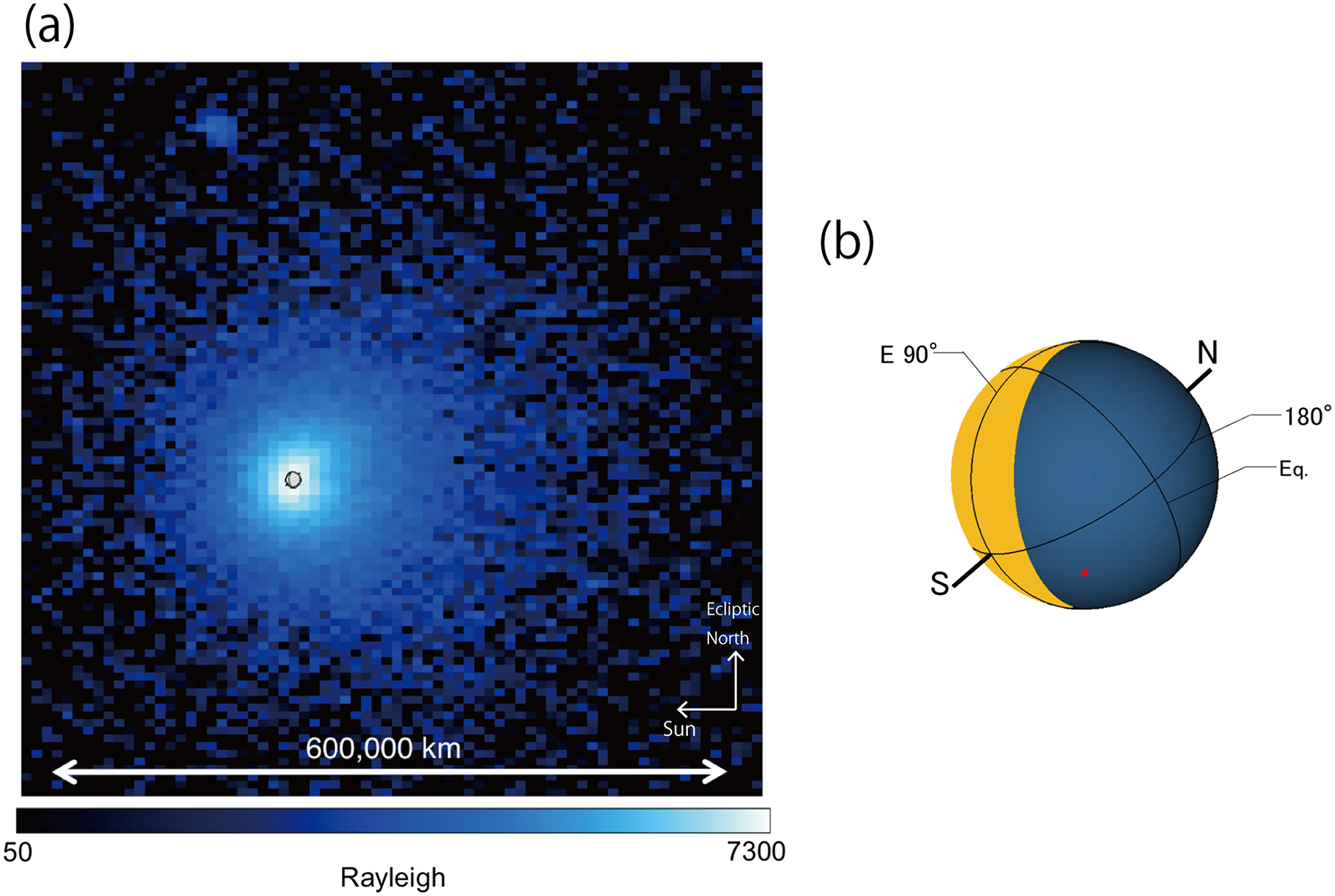}
    \end{tabular}
    \end{center}
    \caption 
    [The Earth's Lyman-$\alpha$ geocorona as observed from a distance of 0.1~au.]{ \label{fig:earth}
    The Earth's Lyman-$\alpha$ geocorona as observed from a distance of 0.1~au (panel a). The Earth's exosphere is predominantly composed of neutral hydrogen, which escapes the lower atmosphere after being photodissociated into H and O atoms \cite{Kameda2017}. Panel b represents the illumination of Earth by the Sun at the time of the observation. HWO may be capable of detecting this feature in Earth-like exoplanets, and we aim to make sure the its UV spectrograph is capable of doing so.} 
    \end{figure}

The Earth has a unique feature among planets in the Solar System: a large, H-rich exosphere that spans more than 38 Earth-radii \cite{Kameda2017} (see Figure \ref{fig:earth}). This cloud is fed by escaping H resulting from photodissociation of water molecules in the Earth's upper atmosphere \cite{Jura2004}. Venus and Mars also possess H exospheres, but are significantly less extended than Earth's \cite{Kulikov2007}. Thus, it has been posed that the search for large, H-rich exospheres around transiting exoplanets provides another pathway for the identification of potentially habitable worlds \cite{Castro2018}. 

We propose to carry out such a search in a handful of the most promising habitable-zone rocky planets around M dwarfs. If one of these possesses a large neutral H fed by atmospheric escape, it will produce an excess absorption in the stellar Lyman-$\alpha$ line as it transits. The depth of this feature is proportional to the size, ionization state and density of the cloud, as well as the stellar host properties (namely radius, radial velocity and distance). Previous attempts at observing this feature around potentially Earth-like exoplanets around TRAPPIST-1 with HST only yielded non-detections \cite{Bourrier2017a, Bourrier2017b}. 

\begin{figure}[t]
    \begin{center}
    \begin{tabular}{c}
    \includegraphics[width=0.75\textwidth]{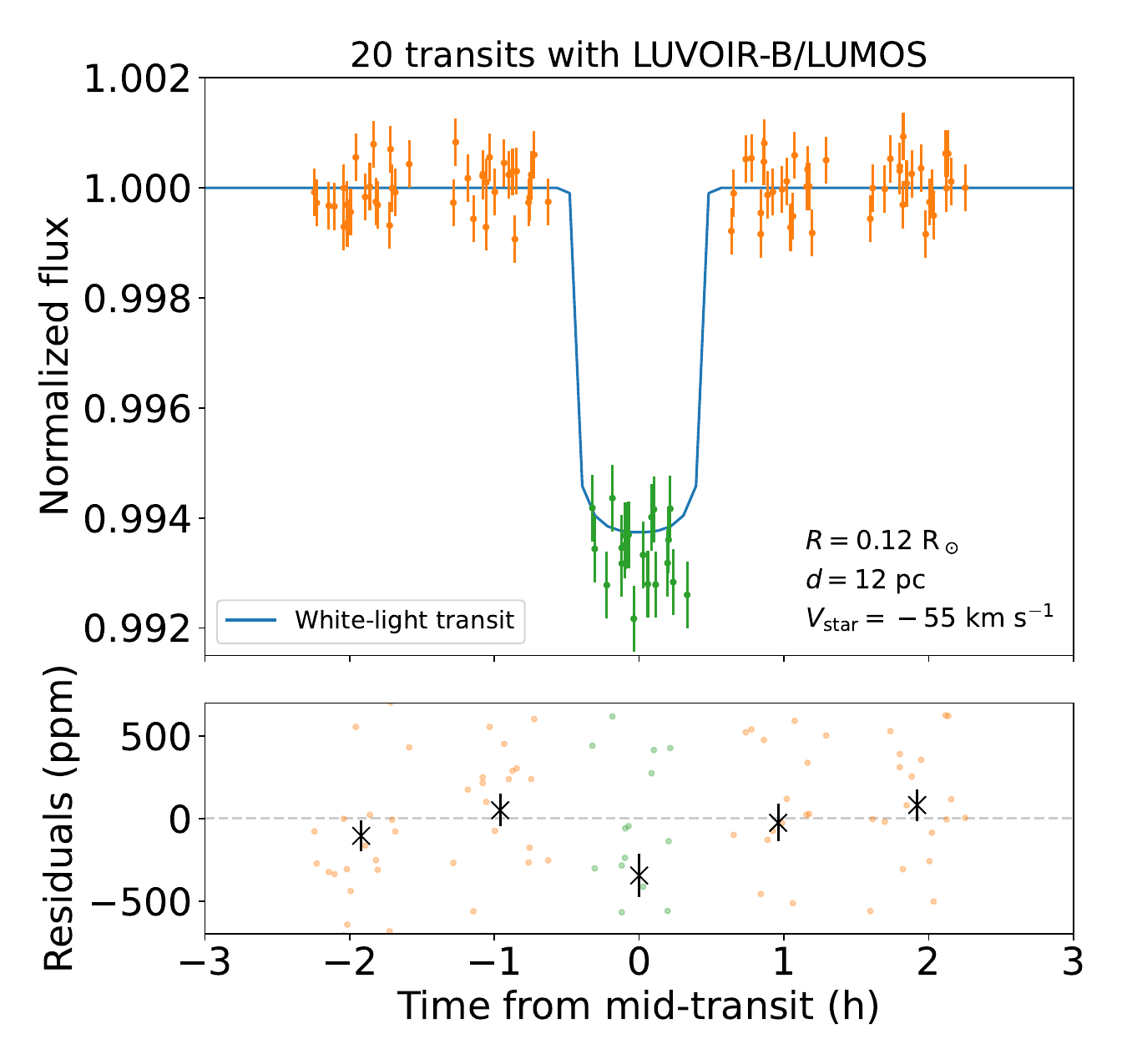}
    \end{tabular}
    \end{center}
    \caption 
    [Simulated Lyman-$\alpha$ transit light curve of an Earth-like planet transiting an M dwarf.]{ \label{fig:earth_lc}
    Simulated Lyman-$\alpha$ transit light curve of an Earth-like planet transiting an M dwarf at a distance of 12 pc and with a radial velocity of $-55$~km\,s$^{-1}$ \cite{DSantos2019a}. The signal of the Earth-like exosphere is seen as an excess absorption of $\sim 500$~ppm (green data points) in the residuals after subtracting the transit depth of the planet's opaque disk. This signal can be detected at 4-sigma confidence within 20 transits, assuming the instrumental parameters from the concept studies of \emph{LUVOIR}-B/LUMOS.} 
    \end{figure}

Dos Santos et al. \cite{DSantos2019a} simulated the observable Lyman-$\alpha$ transit of a rocky planet with an Earth-like exosphere around M dwarfs to investigate whether it would be observable with current or future facilities. The reason for the focus on M-dwarf host stars is because they provide better planet-to-star contrast ratios than solar-type hosts, which increases the signal-to-noise ratio of transit observations. Dos Santos et al. took an empirical model of the Earth's exosphere based on an observation from the LAICA/PROCYON instrument \cite{Kameda2017} and calculated, through radiative transfer, the transit depth of such an exosphere around stars at varying distances and physical parameters. They assumed a spectral resolving power $R = 44\,000$ and an instrument's effective area of $73\,100$~cm$^2$. After this exercise, they found that the Earth's exosphere transiting an M dwarf similar to TRAPPIST-1 absorbs up to approximately 500~ppm of the stellar light in Lyman-$\alpha$ (see Fig. \ref{fig:earth_lc}), which is not detectable with \emph{HST}. If we aim to characterize the exospheres of Earth-like exoplanets using \emph{HWO}, we should strive to achieve this level of precision in Lyman-$\alpha$ fluxes. 

In preparation for this program, the community should aim to understand the impact of stellar activity \cite{Hall2008} in high-precision Lyman-$\alpha$ light curves with \emph{HWO}, which was not discussed in Dos Santos et al. \cite{DSantos2019a}. Additionally, more theoretical simulations of exospheres in rocky planets around M dwarfs should be conducted, particularly on how they could differ from ours if we assume an Earth-like lower atmosphere irradiated by an M dwarf spectrum \cite{Segura2010, Badhan2019}.

\subsubsection{A survey of atmospheric escape in a large sample of exoplanets}

\begin{figure}[t]
    \begin{center}
    \begin{tabular}{c}
    \includegraphics[width=0.95\textwidth]{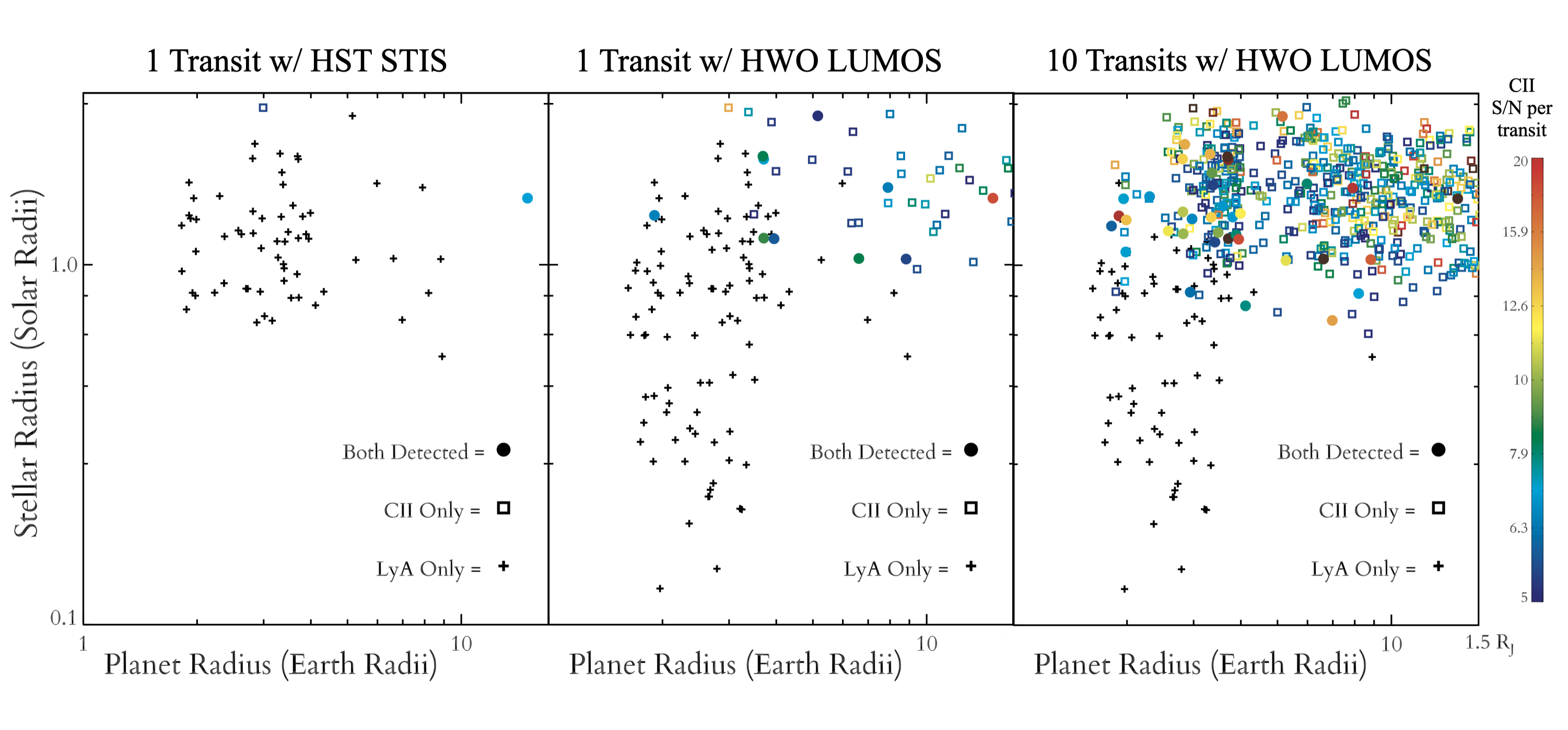}
    \end{tabular}
    \end{center}
    \caption 
    [Simulated population of exoplanets to be observed in the UV with \emph{HWO}.]{ \label{fig:sample}
    Simulated population of exoplanets that can have significant detections of escaping hydrogen and/or ionized carbon if observed with \emph{HST}/STIS and \emph{HWO}/LUMOS in a given number of transits. By co-adding several transits with HWO it will be possible to fully characterize escape from dozens of exoplanets across parameter space.} 
    \end{figure}

In addition to the deep campaign explained above, we advocate for a broad survey of atmospheric escape signatures in a large sample of transiting exoplanets within a range of planetary properties, varying from hot Jupiters to cool sub-Neptunes \cite{Lopez2019}. As an example for a potential yield, we take the simulated TESS planet catalog from Barclay et al.\cite{Barclay2018}, combine it with the mass-radius relation from Wolfgang et al. \cite{Wolfgang2016} and the escape models from Lopez \cite{Lopez2017}. Using those, we predicted escape rates for solar composition H$_2$-dominated atmospheres for each planet in the TESS catalog and then estimated the transit signal-to-noise ratio in both Lyman-$\alpha$ and C\,{\sc ii} scaled from the GJ~436b observation \cite{Ehrenreich2015}. Lastly, we applied a distance cut off at 50~pc for Lyman-$\alpha$ observations due to interstellar medium extinction. For now, one can adopt 50 to be a representative number of planets that can be feasibly observed (see Fig. \ref{fig:sample}). However, in the future, the exact sample size and the strategy should be thoroughly investigated in order to maximime the information content of the observations, such as the recent investigations carried out for \emph{JWST} observations \cite{Batalha2023, Ih2025}.

\begin{figure}[t]
    \begin{center}
    \begin{tabular}{c}
    \includegraphics[width=0.75\textwidth]{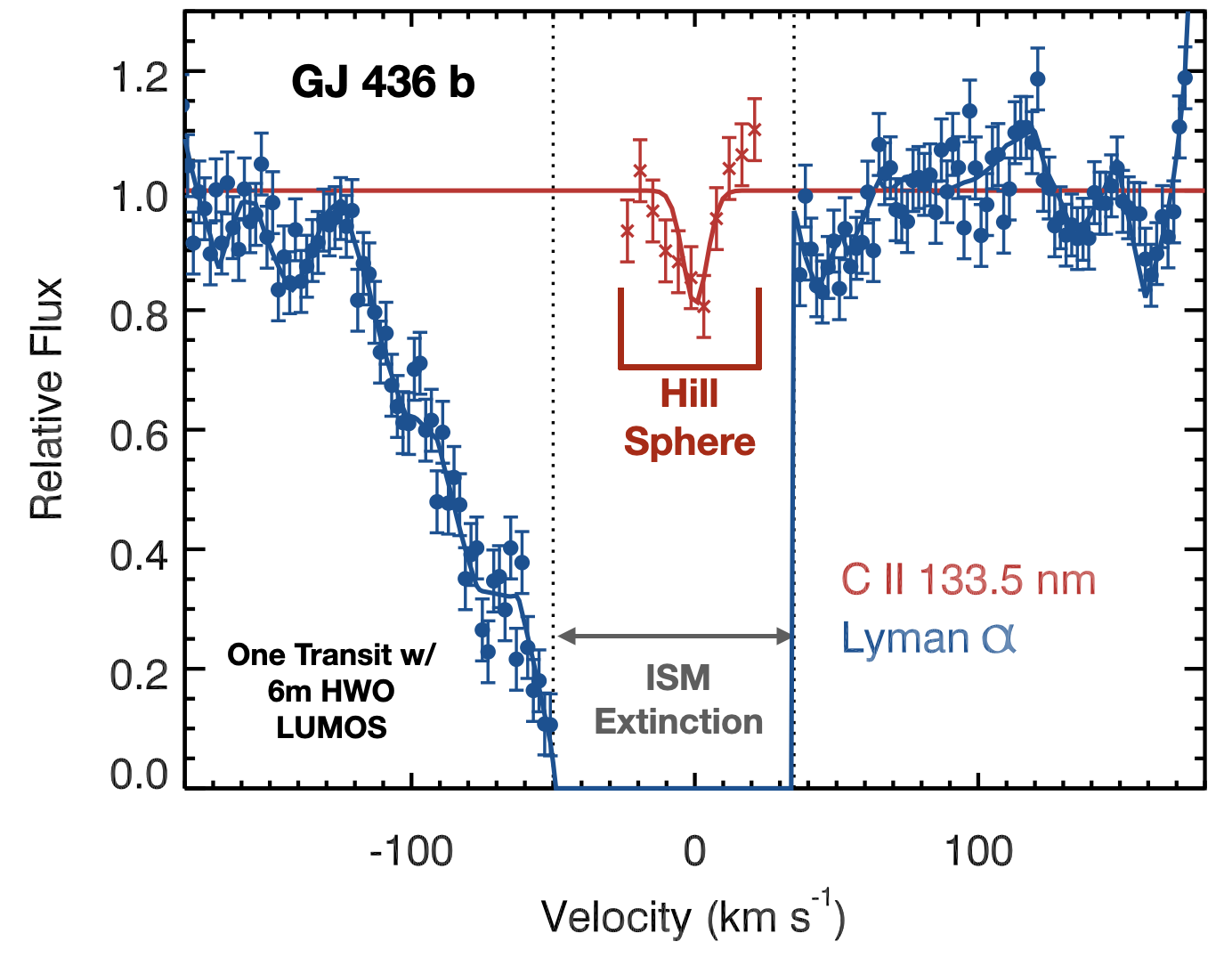}
    \end{tabular}
    \end{center}
    \caption 
    [Simulated UV transmission spectrum of GJ~436b with \emph{HWO}.]{ \label{fig:gj436b}
    Simulated UV transmission spectrum of the Neptune-sized exoplanet GJ~436~b in the Lyman-$\alpha$ line (which traces escaping hydrogen; blue) and singly-ionized carbon (red). These spectra are shown in Doppler velocity space (x-axis) instead of wavelength space, and null velocity corresponds to the rest wavelength of the spectral line in the stellar rest frame. This simulation assumes one transit with \emph{HWO}, with a 6-m telescope aperture and the original characteristics of the LUMOS spectrograph. Figure is adapted from LUVOIR STDT; E. Lopez (NASA/GSFC) \& K. France (Univ. Colorado).} 
    \end{figure}

Building on past efforts that studied transits in Lyman-$\alpha$ and C\,{\sc ii}, we simulated how these two signatures would present themselves in one transit with \emph{HWO} (see the example of the warm Neptune GJ 436b in Fig. \ref{fig:gj436b}). These represent only two of the signals that can be detected but, as seen in Linssen \& Oklop{\v{c}}i{\'c} \cite{Linssen2023}, there is a wide range of other spectral tracers available in the UV that should be included in the wavelength access of \emph{HWO} and its UV instrument.

While several of these lines have been detected at low signal-to-noise ratio (S/N) with \emph{HST} for a handful of hot Jupiters \cite{DSantos2023a}, detecting these lines at high S/N is an important for three vital reasons. Firstly, unlike Lyman-$\alpha$, these lines experience relatively little extinction by the interstellar medium, allowing us to directly probe wind-launching conditions, where temperatures are higher and velocities are lower. Secondly, each of these lines probes a slightly different gas temperature allowing us to break observational degeneracies between gas temperature, density, and ionization state \cite{Linssen2023}, which will allow us to fully map the structures of their upper atmospheres and provide crucial constraints on models of atmospheric escape. Lastly, this offers a unique opportunity to directly measure the individual abundances of key metal species in a region of the atmospheres that, unlike optical and near-infrared transmission spectra, cannot by hidden by planetary clouds or hazes.

\section{Conclusions}

The \emph{Hubble Space Telescope} was instrumental for the first observations of exoplanet atmospheres andit still continues to deliver high-quality data that help us understand how planetary systems evolve through atmospheric escape. However, by the decade of 2030's, \emph{HST} is predicted to cease operations and the astronomical community will lose their only publicly-accessible telescope capable of moderate- to high-resolution UV spectroscopy. This capability is essential for us to continue investigations on exoplanet science and general astrophysics.

The next NASA Flagship Mission, aptly named the \emph{Habitable Worlds Observatory}, will be the successor of \emph{HST} and is expected to enable us to do transformative science after its launch by the early 2040's. \emph{HWO} is expected to house a significantly more sensitive UV spectrograph than those currently available on \emph{HST}, and ensuring that it will have the right technical characteristics will be crucial to advance our knowledge of atmospheric escape and planetary evolution.

In particular, we advocate for the development of a UV spectrograph that is capable of: i) moderate- to high-resolution observations, namely with a resolving power better than COS ($R > 10\,000$); ii) covering wavelength ranges between $1\,000 - 2\,000$~\AA; iii) observing high count rates at a spectrophotometric precision $>100 \times$ better than that of the COS instrument; and iv) uninterrupted time-series observations in timescales of several hours.

Such an instrument would enable us to do a deep search for Earth-like exospheres in rocky planets that transit low-mass stars (mainly M-type) using Lyman-$\alpha$ transmission spectroscopy, which is achievable within a reasonable ($\sim$20) number of transits. Furthermore, it would allow us to conduct a broad survey of escaping H and metals in a large sample of transiting exoplanets, which will help us better understand the chemistry and structure of upper atmospheres, as well as how they evolve with time.

% \disclosures 
\subsection*{Disclosures}
The authors declare that there are no financial interests, commercial affiliations, or other potential conflicts of interest that could have influenced the objectivity of this research or the writing of this paper.

\subsection* {Code, Data, and Materials Availability} 
The findings on Earth-like exospheres were published in Dos Santos et al. \cite{DSantos2019a} and the data and codes necessary to replicate those results are in said publication. The data and code related to the simulation of the broad survey of atmospheric escape is available upon request to the authors. The {\tt p-winds} code to simulate isothermal Parker wind models is openly available in \url{https://github.com/ladsantos/p-winds}.

\subsection* {Acknowledgments}
The authors thank the valuable feedback of the community when discussing the atmospheric escape science case for \emph{HWO} in the context of the ``Solar Systems in Context'' working group. L.A.D.S. acknowledges the often-overlooked labor of the custodial, facilities, information technology and security staff at STScI -- this research would not be possible without them. This research is based on observations made with the NASA/ESA Hubble Space Telescope. The data are openly available in the Mikulski Archive for Space Telescopes (MAST), which is maintained by the Space Telescope Science Institute (STScI). STScI is operated by the Association of Universities for Research in Astronomy, Inc. under NASA contract NAS 5-26555. This research made use of the NASA Exoplanet Archive, which is operated by the California Institute of Technology, under contract with the National Aeronautics and Space Administration under the Exoplanet Exploration Program.

%%%%% References %%%%%

\bibliography{report}   % bibliography data in report.bib
\bibliographystyle{spiejour}   % makes bibtex use spiejour.bst

%%%%% Biographies of authors %%%%%

\vspace{2ex}\noindent\textbf{Leonardo A. Dos Santos} is an associate astronomer at the Space Telescope Science Institute and an associate research scientist at Johns Hopkins University. He received his MS degree in astronomy at the Universidade de S\~ao Paulo in 2017 and his PhD degree in astronomy \& astrophysics at the University of Geneva in 2021. He is the author of more than 50 refereed journal articles and made significant contributions to the science operations of the Hubble Space Telescope. Currently, he acts an instrument scientist for the Cosmic Origins Spectrograph. His current research interests include physics of exoplanet atmospheres, UV spectroscopy, IR spectroscopy, and stellar activity.

\vspace{2ex}\noindent\textbf{Eric Lopez} is a research space scientist at the NASA Goddard Space Flight Center. He received his PhD in Astronomy and Astrophysics from the University of California Santa Cruz and has published more than 70 refereed articles. He is also a co-lead of the Exoplanet Modeling and Analysis Center and was involved with studies of the LUVOIR concept. His current research interests include models of planetary structure, composition, evolution, and atmospheric escape.

\listoffigures
\listoftables

\end{spacing}
\end{document}